# Recognition of Emerging Technology Trends.

## Class-selective study of citations in the U.S. Patent Citation Network.*

**Péter Bruck[a], István Réthy[b], Judit Szente[c], Jan Tobochnik[d] and Péter Érdi[e,f,1]**

[a] ProcessExpert Ltd, Budapest, Hungary, [b] Gobinfo Ltd, Budapest, Hungary, [c] Department of Atmospheric, Oceanic and Space Sciences, University of Michigan, Ann Arbor, MI 48109, [d] Departments of Physics and Computer Science, Kalamazoo College, Kalamazoo, MI 49006, [e] Center for Complex Systems Studies, Kalamazoo College, Kalamazoo, MI 49006, and [f] Computational Neuroscience Group, Wigner Research Centre for Physics, Hungarian Academy of Sciences, Budapest

[1] Correspondence to Péter Érdi: Kalamazoo College, 1200 Academy Street, Kalamazoo, MI 49006; +1 269 337 5720; e-mail:perdi@kzoo.edu

**Abstracts** By adopting a citation-based recursive ranking method for patents the evolution of new fields of technology can be traced. Specifically, it is demonstrated that the laser / inkjet printer technology emerged from the recombination of two existing technologies: sequential printing and static image production. The dynamics of the citations coming from the different "precursor" classes illuminates the mechanism of the emergence of new fields and give the possibility to make predictions about future technological development. For the patent network the optimal value of the PageRank damping factor is close to 0.5; the application of d=0.85 leads to unacceptable ranking results.

**Keywords** PageRank · Patents · Trend recognition · Damping factor

# 1 Introduction

We have previously studied the growth of the patent citation network both at the "microscopic" level of individual patents (Csárdi et al. 2009; Érdi 2007; Strandburg et al. 2007; Csárdi et al. 2009; Strandburg et al. 2009) and at "mesoscopic" (Érdi et al. 2013) levels. Microscopic level studies helped to measure the "attractiveness" of a patent, as a function of its age and the number of citations already obtained. At the mesoscopic level the analysis has been extended to subclasses, and it was demonstrated that it is possible to detect and predict an emerging new technological trend through the application of an appropriate clustering algorithm.

The patent citation network can be viewed as a *time-evolving complex system*. Inventions often can be described as combinations of already existing technologies (Schumpeter 1939; Valverde et al. 2007; Henderson et al. 1990; Weitzman 1996; Hargadon et al. 1997). For example, one might think of the automobile as a combination of the "horse carriage and the internal combustion engine" [Fleming 2001; Perra et al. 2008; Podolny et al. 1995; Podolny et al. 1996; Fleming et al. 2001). For the combination a new and relevant illustration will be shown in Section 4.3 of the present paper: it is demonstrated that the emergence of industry trends can be identified by observing the temporal development of the proportion of class-specific citations.

# 2 Recursive ranking: from web pages to patents

In the theory of social networks centrality measures were constructed to rank network nodes based on their topological importance (Weitzman1996). These centrality measures reflect that either there is a connection between a pair of nodes or there isn't, and thus the elements of the adjacency matrix of the graph are zeros or ones. Another family of measures is related to the spectral properties of the adjacency matrix (Perra et al. 2008; Vigna 2009), taking into account the importance of the neighbors.

Brin and Page determined the importance of a node by taking into consideration both the number and the importance of its immediate neighbors. Since the neighbors also have neighbors, the calculation of the matching centrality measure - PageRank (Brin et al. 1998) - has to be recursive. The PageRank (P) values of the nodes are calculated iteratively:

$$P_i^{(t+1)} = (1-d)/N + d\sum_{j=1}^{n_i} \frac{P_j^{(t)}}{m_j} \qquad (1)$$

where $P_i^{(t)}$ is the PageRank value of node *i* at iteration *t*, *N* is the total number of the nodes in the network, $n_i$ is the count of incoming links of node *i*, $m_j$ is the count of outgoing links of node *i* and *d* is called the "damping factor". Each node that is cited in the network by node *j* receives the same share of the PageRank. Equation (1) reflects that not all of the resources of node *i* will be distributed through its outgoing links among the neighbors: a fraction (*1-d*) of its resources is evenly distributed among the N nodes of the network. It also reflects that two independent mechanisms contribute to the distribution of the resources: the contribution of the first one is determined by the number of network nodes while the second one depends primarily on the structure of the network

* The authors benefited from discussions with colleagues and students both in Budapest and Kalamazoo. Special thanks to Gábor Csárdi, János Tóth and Péter Volf for commenting previous versions of the manuscript. PE thanks the Henry Luce Foundation for support of Complex Systems Studies as Henry R Luce Professor. JT thanks the Herbert H. and Grace A. Dow Foundation for support as the Dow Distinguished Professor of the Natural Sciences.
Conflict of Interest: The authors declare that they have no conflict of interest.

graph (i.e. the number of neighbors ). The relative contribution of the two mechanisms is determined by the parameter *d,* – where the smaller is the value of d, the lower is the relative influence of the neighbors in the graph.

If a network has a node which has no outgoing links, (1) cannot be applied, because the value of $m_j$ is zero. Brin et al. (1998) proposed to distribute the PageRank of such "dangling" nodes evenly among all other nodes of the network through new, "artificial" links. The evenness of the redistribution guarantees that the relative ranks of the other nodes remain unchanged.

To calculate PageRank in the first step we attribute the same initial PR value to each node; then we calculate a new set of PR value for each node and this process is repeated until at each node a stationary value is reached. The method of Brin and Page guarantees that the iteration procedure will converge, but a high price has been paid for this advantage: by the introduction of the new links the structure of the original graph is drastically changed.

The spirit of the recursive PageRank algorithm can be extended to patent citation analysis as well. A patent is useful if it contains significant, reusable information. The proof of the usefulness is that the patent is being cited by other patents. The importance of a patent is measured by the frequency of the citation of the given patent by other patents; however, the weight of these citations is not equal: more important are those citations that are cited by important citations.

Citations related to a given patent are either incoming citations made by newer patents (called inlinks), or citations that are contained in the text of the patent itself (called outlinks). Each citation is always an inlink and an outlink at the same time - depending on the viewpoint. The number of outlinks of a patent are fixed at the moment when the patent is granted, while the number of inlinks of a patent keeps growing in time whenever a new citing patent appears. To calculate the centrality of a given node we may take into account its inlinks, its outlinks or both. The present work will focus on inlinks, because inlinks reflect the progress of technology. Due to the nature of the patent system inlink citations are strictly unidirectional, acyclic and singly connected. Due to these restrictions the structure of the inlink graph significantly differs from the structure of the web graph.

Shaffer (2011) - motivated by the PageRank algorithm - introduced Patent Rank as a centrality measure of patents registered by the U.S. Patent and Trademark Office (USPTO). His primary aim was to assign a single value to each patent, which simultaneously reflects both the relative significance of a patent as well as its economic value. Similarly to PageRank, Patent Rank also applies a recursive algorithm, and utilizes inlinks as well as outlinks. To avoid the problem caused by the zero elements of the adjacency matrix due to nodes without incoming or outgoing links Shaffer applied the matrix augmentation technique: the U.S. Patent Office is used as a super-node which is cited by all patents in the network and cites all patents in the network. This is the point where the structure of the original graph is being changed. His approach to ensure irreducibility using matrix augmentation appears to be less intrusive than the solution of PageRank, where every node becomes directly connected to every other node, but it was shown (Tomlin 2003; Langville et al. 2003) that the two methods are mathematically equivalent; however, the question, as to how to choose the damping factor for this equivalence was left open.

Shaffer used his method to rank the patents of the USPTO database. The top 20 patents of Shaffer are shown in the last column of Table 2.

# 3 Methods

## 3.1 Database

To study the temporal evolution of the USPTO database we apply the original PageRank centrality measure, since - contrary to the augmentation technique - PageRank has an additional control parameter: the damping factor (*d*), which enables the fine tuning of the results.

We developed a PL/SQL program to calculate the PageRank value of each node using the original iterative algorithm of Brin et al. (1998)[†]. The correctness of calculated PageRank values of our software implementation were tested on two classic examples from the Netlogo simulator (Stonedahl et al. 2009). From the digital data of all patents granted in the 1976-2012 period by the USPTO (approximately 3.94 million patents) we constructed an Oracle repository. From the 44 million inlink citations we generated five complete PageRank sets corresponding to the full range of d values: 0.01, 0.15, 0.50. 0.85 and 0.99.

The time requirement of one iteration using an Oracle VM under Win7 was approx. three minutes. Iterations were terminated when the sum of the differences between the new PageRank values and the previous ones for the 3.94 million nodes became smaller than $10^{-6}$. The number of iterations required to reach the equilibrium value of PageRank was low between 4 ( d = 0.01) and 18 ( d = 0.99).

[†]The source code is available under Supporting Information (Section 1).

### 3.2 The optimal values of d for citation networks

No objective criteria have been published to support the selection of the optimal value of *d* . To achieve fast convergence Brin and Page originally proposed to use *d* = 0 .85 value for the web graph, and this value has been applied to a broad range of networks. However, when Avrachenkov et al. (2008) examined the ergodic structure and the probability flow of the web graph, it was recognized that with *d* = 0 .85 the most valuable central part of the network does not receive its fair share of the PageRank flow, since the peripheries of the network absorb too much PageRank; they suggested that *d*=0.50 would be a more appropriate choice. For a network of scientific citations also *d* = 0.5 has been selected [Chen et al. 2007; Maslov 2009; Walker et al. 2007), but the choice was intuitive: researchers hardly go deeper than two levels when references in cited papers are checked.

# 4 Results

[Table 1 about here.]

The most relevant results of the calculations are summarized in Table 1 and Table 2.[‡] Table 1 displays the Number of Citations (NCIT) and the corresponding PageRank values of the top 20 patents. We also indicated which citations originated from the two most important patent classes: Class 435 - related to genetics - is marked by green, while Class 347 - corresponding to inkjet and laser printer technology - is marked by yellow. This coloring can provide a visual impression showing which part of the patent graph receives preference at the selected value of *d*.

### 4.1 Selection of the optimal value of d for the present study

We will demonstrate that the proper choice of d for the patent network is in the 0.01 - 0.50 range. This unusual result contradicts most experiences gained with various subsets of the web graph where the selected value of *d* usually is 0.85.

Let us summarize what happens when the value of *d* is changed in the [0,1) interval.

• If *d* = 0, the PageRank of each and every node would be the same: 1 /N - this choice is unsuitable for ranking; the redistribution is solely controlled by the initial uniform distribution and the neighbors have no effect whatsoever.

• If *d* = 0.01 (Table 1 Column 1), it is generally expected (Fortunato et al. 2007) that for very low *d* values PageRank will be determined by the above uniform distribution, because the effect of the neighbors through the internal link structure is heavily dampened. Our results - shown in the first column of Table 1 - contradict this expectation. Though the influence of the neighbors is heavily dampened, the information infiltrates quickly: even in the case of the large USPTO network four iterations are sufficient to reach equilibrium and the result list seems to be absolutely relevant. This range of *d* was not studied earlier in detail; our results show that it deserves more attention.

• If *d* is in the 0.01 to 0.50 range (Table 1, Column 2 and 3), the characteristics of top ranking patents look well balanced. Though they are well cited, an acceptable amount of lower cited patents could make it to the list - thanks to their important neighbors. This indicates that the weight attributed to the central part of the network and to the peripheries is realistic. It is reasonable to assume, that the centrality measure will express important characteristics of the patent network, if the *d* values are selected from this range. Within this range the ranking is remarkably stable (Ding et al. 2009), and the results are similar to the results of Shaffer (Table 1, last column).

• If *d* is in the 0.50 to 0.99 range (Table 1, Columns 3 and 4), the most important central components of the graph do not receive their fair share of the PageRank mass, as was recognized by Boldi et al (2005, 2015). In the early days of PageRank applications it was supposed that choosing *d* close to unity would lead to 'truer' PageRank values. Recent investigations proved that this supposition is not correct: for real-world graphs such *d* values do not result in meaningful ranking: PageRank becomes concentrated in buckets having no connection with other network components and the PageRank of the well connected core part of the network graph becomes heavily underrated. The consequence: for *d* = 0.99 the list of the top 20 patents is full of irrelevant results.

The most interesting result of Table 1 is, that in the case of the patent network - contrary to the web graph - *d* = 0.85 is a bad choice.

These results are in good correspondence with Avrachenkov et al (2008) and Boldi et al. (2005); this led us to choose *d* = 0.5 for the present investigations.

---

[‡] All results of the top 100 ranked patents for each value of *d* are available under Supporting Information (Section 2).

| | PageRank (1976-2012) | | | | | | | | | | | | | | | Shaffer (1976-2000) | | |
|---|---|---|---|---|---|---|---|---|---|---|---|---|---|---|---|---|---|---|
| | NCIT | PR (d=0.01) | PatentNo | NCIT | PR (d=0.15) | PatentNo | NCIT | PR (d=0.5) | PatentNo | NCIT | PR (d=0.85) | PatentNo | NCIT | PR (d=0.99) | PatentNo | NCIT | Patent Rank | PatentNo |
| | | iterations:4 | | | iterations:7 | | | iterations:12 | | | iterations:16 | | | iterations:18 | | | | |
| 1 | 2361 | 9,3639E-07 | 4683195 | 2361 | 1,1640E-05 | 4683195 | 2361 | 4,2023E-05 | 4683195 | 60 | 1,5500E-04 | 3988545 | 60 | 3,2393E-04 | 3988545 | 2256 | 2,1700E+05 | 4683202 |
| 2 | 2680 | 8,8705E-07 | 4683202 | 2680 | 1,0923E-05 | 4683202 | 2680 | 4,0540E-05 | 4683202 | 292 | 1,2169E-04 | 4237224 | 48 | 1,6948E-04 | 4074232 | 2018 | 2,0900E+05 | 4683195 |
| 3 | 1765 | 8,8354E-07 | 5523520 | 1765 | 9,2834E-06 | 5523520 | 292 | 3,9499E-05 | 4237224 | 62 | 9,0660E-05 | 3932805 | 62 | 1,6806E-04 | 3932805 | 286 | 1,5100E+05 | 4237224 |
| 4 | 442 | 8,0959E-07 | 5536637 | 442 | 8,0158E-06 | 5536637 | 73 | 2,5677E-05 | 4395486 | 48 | 8,9487E-05 | 4074232 | 292 | 1,6227E-04 | 4237224 | 558 | 1,3000E+05 | 3702886 |
| 5 | 2016 | 6,1096E-07 | 4723129 | 2016 | 6,2606E-06 | 4723129 | 1765 | 2,3075E-05 | 5523520 | 202 | 7,5959E-05 | 3976982 | 202 | 1,3263E-04 | 3976982 | 297 | 1,2300E+05 | 3747120 |
| 6 | 891 | 5,4436E-07 | 5367109 | 180 | 5,2072E-06 | 4812599 | 2016 | 2,1376E-05 | 4723129 | 365 | 7,2709E-05 | 3956615 | 365 | 1,2971E-04 | 3956615 | 427 | 1,1700E+05 | 4358535 |
| 7 | 885 | 5,4108E-07 | 5850009 | 1731 | 5,0339E-06 | 4463359 | 754 | 2,1234E-05 | 4558413 | 27 | 7,1177E-05 | 4309756 | 139 | 1,1429E-04 | 3950733 | 206 | 1,1000E+05 | 3856513 |
| 8 | 875 | 5,3275E-07 | 5304719 | 1714 | 4,8821E-06 | 4740796 | 151 | 1,9646E-05 | 4251824 | 139 | 7,0605E-05 | 3950733 | 59 | 1,1313E-04 | 3934122 | 1955 | 1,0300E+05 | 4723129 |
| 9 | 180 | 5,2913E-07 | 4812599 | 292 | 4,8115E-06 | 4237224 | 467 | 1,9060E-05 | 4358535 | 73 | 6,9164E-05 | 4395486 | 27 | 1,1252E-04 | 4309756 | 1691 | 9,8000E+04 | 4463359 |
| 10 | 1714 | 5,2417E-07 | 4740796 | 891 | 4,6031E-06 | 5367109 | 1731 | 1,8863E-05 | 4463359 | 754 | 6,8317E-05 | 4558413 | 44 | 1,0783E-04 | 3984626 | 91 | 8,9000E+04 | 3950357 |
| 11 | 1731 | 5,2350E-07 | 4463359 | 885 | 4,5206E-06 | 5850009 | 180 | 1,8771E-05 | 4812599 | 59 | 6,6249E-05 | 3934122 | 291 | 1,0415E-04 | 4063220 | 70 | 8,8000E+04 | 4395486 |
| 12 | 1581 | 4,9496E-07 | 4345262 | 1606 | 4,4581E-06 | 4558333 | 442 | 1,8438E-05 | 5536637 | 2361 | 6,5568E-05 | 4683195 | 63 | 9,7086E-05 | 3984822 | 136 | 8,6000E+04 | 4172124 |
| 13 | 1606 | 4,9484E-07 | 4558333 | 1581 | 4,4493E-06 | 4345262 | 1376 | 1,6792E-05 | 5572643 | 2680 | 6,5408E-05 | 4683202 | 754 | 9,4636E-05 | 4558413 | 168 | 8,6000E+04 | 5045417 |
| 14 | 1248 | 4,8905E-07 | 4816567 | 875 | 4,3971E-06 | 5304719 | 588 | 1,6506E-05 | 4539507 | 291 | 6,2525E-05 | 4063220 | 56 | 9,4520E-05 | 3938096 | 463 | 8,6000E+04 | 4367924 |
| 15 | 1546 | 4,8612E-07 | 4313124 | 1546 | 4,3299E-06 | 4313124 | 1714 | 1,6488E-05 | 4740796 | 98 | 6,2408E-05 | 4298685 | 103 | 9,3082E-05 | 3937925 | 143 | 8,5000E+04 | 3778614 |
| 16 | 1507 | 4,7254E-07 | 4459600 | 1248 | 4,2162E-06 | 4816567 | 1441 | 1,6418E-05 | 5103459 | 44 | 5,8364E-05 | 3984626 | 68 | 9,0401E-05 | 3996657 | 619 | 8,5000E+04 | 4558413 |
| 17 | 982 | 4,4244E-07 | 4965188 | 1507 | 4,0978E-06 | 4459600 | 60 | 1,6005E-05 | 3988545 | 69 | 5,7071E-05 | 4164745 | 98 | 8,9454E-05 | 4298685 | 1656 | 8,4000E+04 | 4740796 |
| 18 | 896 | 4,3660E-07 | 4946778 | 467 | 3,7709E-06 | 4358535 | 504 | 1,5362E-05 | 4356429 | 151 | 5,5841E-05 | 4251824 | 31 | 8,9395E-05 | 3968478 | 16 | 8,3000E+04 | 3813316 |
| 19 | 1441 | 4,2883E-07 | 5103459 | 1441 | 3,7200E-06 | 5103459 | 1606 | 1,5306E-05 | 4558333 | 56 | 5,3928E-05 | 3938096 | 73 | 8,6045E-05 | 4395486 | 152 | 8,3000E+04 | 3859527 |
| 20 | 344 | 4,2686E-07 | 5801154 | 982 | 3,4860E-06 | 4965188 | 1581 | 1,5187E-05 | 4345262 | 58 | 5,2307E-05 | 3984822 | 55 | 8,5346E-05 | 4144412 | 1550 | 8,1000E+04 | 4345262 |
| | | | | | | | | | | | | | | | | | | |
| | ∑=26691 | | | ∑=26210 | | | ∑=21782 | | | ∑=7889 | | | ∑=2890 | | | ∑=14758 | | |

**Table 1. PageRank of top 20 USPTO Patents as the function of the damping factor**

Column 6: Data are cited for comparison from Shaffer, Monte: Entrepreneurial Innovation: Patent Rank and Marketing Science. PhD Dissertation, p 213, Washington State University (2011) https://research.wsulibs.wsu.edu/xmlui/bitstream/handle/2376/2902/Shaffer_wsu_0251E_10146.pdf?sequence=1

## 4.2 Patent classes of top ranking patents in the USPTO database

In the list of the twenty patents having the highest PageRank (Table 2) four of the top patents are related to genetics (Class 435). Our ranking method seems to be justified, because the inventor of the two highest ranked patent, Kary B. Mullis, received the 1993 Nobel Prize in Chemistry for his essential contribution to the development of DNA based chemistry (Mullis 1993).

| | Patent Nr | Patent Class | Patent Title | Number of Citations | PageRank * $10^8$ | | | | |
|---|---|---|---|---|---|---|---|---|---|
| | | | | | d=0.50 | d=0.01 | d=0.15 | d=0.85 | d=0.99 |
| 1 | 4683195 | 435 | Process for amplifying, detecting, and/or-cloning nucleic acid sequences | 2361 | 4202 | 94 | 1164 | 6556 | 6927 |
| 2 | 4683202 | 435 | Process for amplifying nucleic acid sequences | 2680 | 4054 | 88 | 1092 | 6540 | 7002 |
| 3 | 4237224 | 435 | Process for producing biologically functional molecular chimeras | 292 | 3950 | 42 | 481 | 12169 | 16227 |
| 4 | 4395486 | 435 | Method for the direct analysis of sickle cell anemia | 73 | 2568 | <33 | 247 | 6916 | 8604 |
| 5 | 5523520 | 800 | Mutant dwarfism gene of petunia | 1765 | 2307 | 88 | 928 | <3000 | <5000 |
| 6 | 4723129 | 347 | Bubble jet recording method and apparatus in which a heating element generates bubbles | 2016 | 2138 | 61 | 626 | 3222 | <5000 |
| 7 | 4558413 | 717 | Software version management system | 754 | 2123 | 35 | 290 | 6832 | 9463 |
| 8 | 4251824 | 347 | Liquid jet recording method with variable thermal viscosity modulation | 151 | 1965 | <33 | 219 | 5584 | 7260 |
| 9 | 4358535 | 435 | Specific DNA probes in diagnostic microbiology | 467 | 1906 | 42 | 377 | 4236 | 5121 |
| 10 | 4463359 | 347 | Droplet generating method and apparatus thereof | 1731 | 1886 | 52 | 503 | 3217 | <5000 |
| 11 | 4812599 | 800 | Inbred corn line PHV78 | 180 | 1877 | 53 | 520 | <3000 | <5000 |
| 12 | 5536637 | 435 | Method of screening for cDNA encoding novel secreted mammalian proteins in yeast | 442 | 1844 | 81 | 801 | <3000 | <5000 |
| 13 | 5572643 | 709 | Web browser with dynamic display of information objects during linking | 1376 | 1679 | 38 | 319 | 3499 | <5000 |
| 14 | 4539507 | 313 | Organic electroluminescent devices having improved power conversion efficiencies | 588 | 1651 | 37 | 311 | 4053 | 5099 |
| 15 | 4740796 | 347 | Bubble jet recording method and apparatus in which a heating element generates bubbles | 1714 | 1649 | 52 | 488 | <3000 | <5000 |
| 16 | 5103459 | 370 | System and method for generating signal waveforms in a CDMA cellular telephone | 1441 | 1642 | 43 | 372 | 3280 | <5000 |
| 17 | 3988545 | 370 | Method of transmitting information and multiplexing device for executing the method | 60 | 1600 | <33 | <192 | 15500 | 32393 |
| 18 | 4356429 | 313 | Organic electroluminescent cell | 504 | 1536 | 35 | 261 | 4251 | 5599 |
| 19 | 4558333 | 347 | Liquid jet recording head | 1606 | 1531 | 49 | 445 | <3000 | <5000 |
| 20 | 4345262 | 347 | Ink jet recording method | 1581 | 1519 | 49 | 444 | <3000 | <5000 |

Table 2. Data of the top 20 USPTO patents, ranked by descending PageRank values (d=0.50)

Over 2000 later patents benefited from the information, transferred from each of these outstanding patents (#4683195 and #4683202) toward the citing patents. But why is patent 4683195 ranked higher than 4683202 if the latter received more citations? In the course of the years the patents which cited the first one proved to be a more valuable resource of information than those patents which cited the second one. In other words: the accumulated role of 4683195 became more important as a resource of information. It is essential for the rest of the paper to understand the quantitative explanation: the PageRank of 4683195 is determined by to the sum of the PageRanks of its citing patents - the higher is this sum, the higher is the amount of information received by the citing documents.

## 4.3 Technological evolution as the recombination of existing technologies.

### 4.3.1 The evolution of laser / inkjet printing

In the top 20 list six of the patents are related to laser / inkjet printing (Class 347). In the past two decades this class became one of the most important USPTO classes: ten percent of the most cited 100 patents in the database belong to Class 347. The citation activity to Class 347 has shown a steady increase between 1976 and 2011; however, behind this deceptively simple tendency a complex picture is hidden.

[Figure 1 about here]

Figure 1 shows the distribution of the PageRank contribution of external patent classes citing Class 347 as the function of time from the early eighties to the end of 2012. The two external classes, which cite Class 347 at the highest frequency are Class 400 (sequential printing mechanism) and Class 358 (static image production).

Because citations are not equal in value, the true contribution of a selected class in a given year cannot be characterized simply by the number of citations originating from this class. Instead, we should use the sum of the PageRanks of the citing patents in that year to characterize the flow of information.

The proportion of the citations originating from Class 358 and Class 400 shows a significant trend change in time: before 2004 the flow toward Class 358 and Class 400 were more or less equal, but after 2004 the contribution connected with Class 400 gradually decreased year by year, while the contribution of Class 358 gradually increased and became dominant in the past years.

### 4.3.2 Class-level trends

The upper chart of Figure 1 reveals an important industry trend: between 1992 and 2012 the interaction between Class 347 and Class 400 contributed to the development of the mechanical construction of the laser / inkjet printers. Citations from Class 358 to Class 400 reflect the contributions of Class 358 to the development of the image formation techniques. Between 1992 and 2004 the contribution of these two classes has been roughly equal, but this trend was not continued after 2004. By 2004 the refinement of printing mechanisms reached a level where further investment apparently became unjustified - the period of mechanical evolution was over. Between 2004 and 2012 the information flow from Class 347 toward Class 358 gradually became significantly larger than the flow toward Class 400. This indicates that the efforts during this period became focused on the development of the imaging part of the solution: reduce the size of droplets, increase the stability of the printed images, etc.

### 4.3.3 Individual-level trends

To check the validity of the above technical trend shown by the class level data let us focus our attention on the six most cited individual patents in Class 347. The question is whether the trend shown by the citation data of the most cited patents have similar trends as the entire class or not. In the pre-2004 period the number of citations from Class 400 was 30 to 40 percent of Class 358 citations. After 2004 it decreased to 8-10 percent and the proportion of Class 358 citations became six times higher. We may conclude that the top six patents follow the same trend as was shown by the whole Class 347.

### 4.3.4 The effect of self-citations

The full text of the top patents in Class 347 revealed two important facts: all of the top patents are granted to Canon and the great majority of the citations to these Canon patents are self-citations by other Canon patents. None of the Canon self-citations was spam - these are genuinely relevant, high quality patents that make rational, technically justifiable citations. Canon has a well-designed and consistently executed patent writing policy: always cite all related Canon patents. This helps to bring the excellent patents of Canon to the top of the list and at the same time prevents the patents of other producers from occupying any of the top positions. Self citation is a common practice used in scientific articles and provides valuable information to the reader. However, it should be recognized that in the case of patents self-citations do not indicate real information flow among the parties; therefore the weight to be attributed to self citations should be smaller.

How to ensure that the sharp change after 2004 - which was originally interpreted as a general industry trend - is not an artifact, caused by the internal decision of a single, very powerful producer? To find out, whether the citation data of the competitors show a similar change, we created a subset of the data where Canon-related patents are not taken into account: patents which cite Canon patents as well as patents which are cited by Canon patents were disregarded (Figure 1, lower chart). Using this reduced dataset the trend proved to be the same as in the case of the complete database (which included Canon patents).

# 5 Conclusions

We calculated the PageRank values of 1976-2012 USPTO patents for five values of the damping factor. For large www networks traditionally $d$=0.85 has been used, but in the case of the patent citation network this value of the damping factor leads to unacceptable ranking .As Avrachenkov predicted, the optimal value of $d$ for this network is close to 0.5.

The class-selective study of the temporal evolution of the citations of Class 347 has revealed certain general industry trends: the two most influential external classes which cite Class 347 at outstanding frequency are Class 400 (sequential printing mechanism) and Class 358 (static image production). The laser / inkjet printer technology seems to emerge from the recombination of these two - already existing - technologies.

The class-selective analysis revealed that the refinement of the printing mechanisms reached such a level by 2004 where further major investment became unjustified. Therefore from the middle of the decade the technical efforts were primarily focused on the development of the imaging part of the solution. This industry trend is equally supported by the study of the patents of all manufacturers in Class 347, as well as the subset of non-Canon patents, and has been verified by the analysis of the most important individual patents in this class. We demonstrated using the example of an important patent class that the evolution of technology trends can be successfully recognized through the study of class-specific change of citation activity in time.

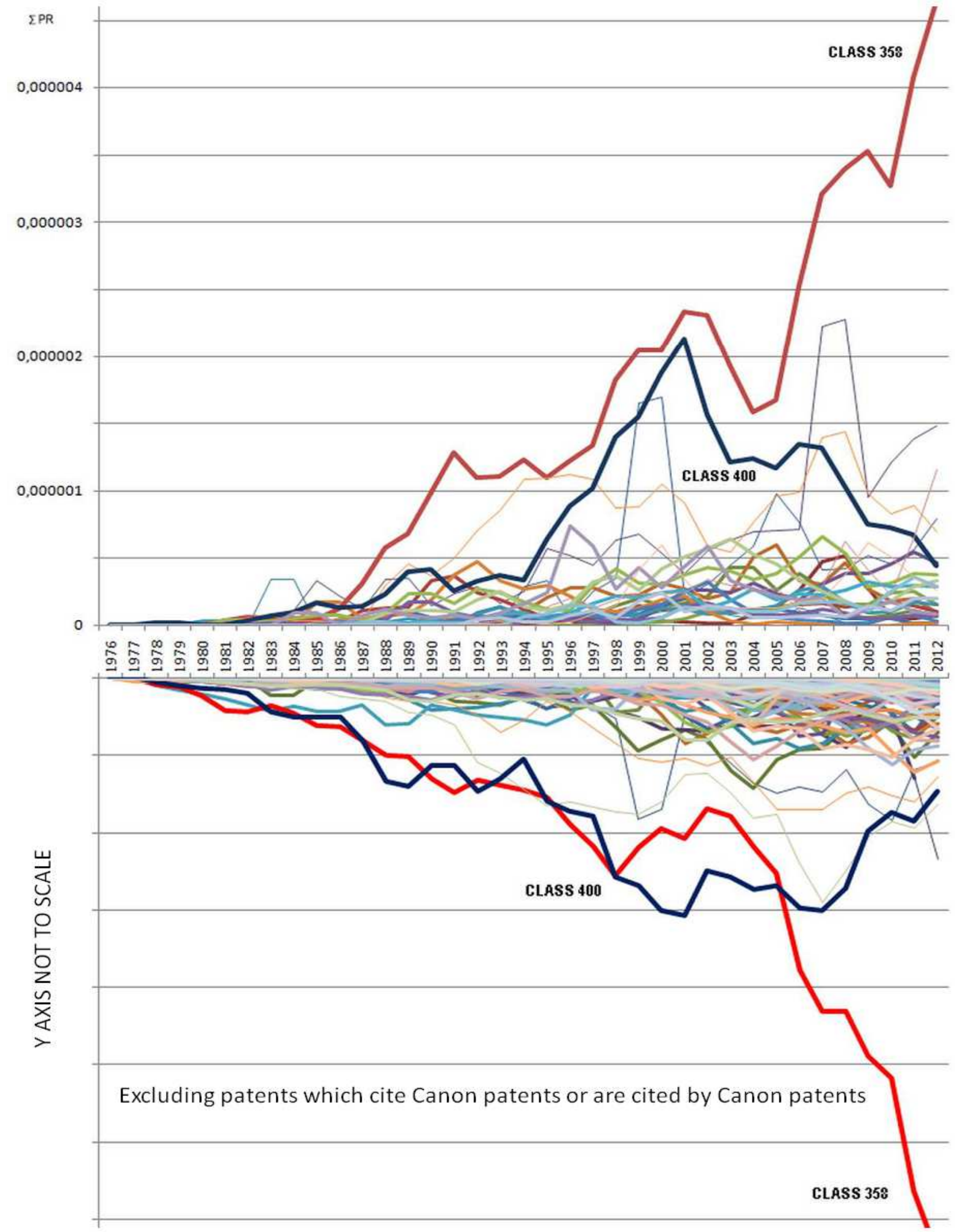

**Figure 1 The time distribution of the PageRank contribution of external patent classes citing Class 347**

The upper chart corresponds to the complete USPTO database; the lower chart shows only non-Canon related data